\documentclass[twocolumn,prd,nofootinbib,aps,prl,floats,floatfix,amsmath,amssymb,longbibliography,secnumarab
ic,preprintnumbers]{revtex4-2} %
\usepackage[final]{graphicx}

\usepackage{hyperref}
\usepackage{amsmath}
\usepackage{bbm}
\usepackage{bm}
\usepackage{amsfonts}
\usepackage{amssymb}
\usepackage{latexsym}
\usepackage{mathtools}
\usepackage{graphicx}
\usepackage[english]{babel}
\usepackage{multirow}
\usepackage{float}
\usepackage{url}
\usepackage{slashed}
\usepackage{xcolor} 
\usepackage[utf8]{inputenc}
\usepackage{stmaryrd} 
\usepackage{enumitem}
\usepackage{hyperref}
\usepackage{cleveref}
\usepackage{siunitx}
\usepackage{verbatim}
\usepackage{relsize}
\usepackage{ulem}

\newcommand{\be}{\begin{equation}}
	\newcommand{\ee}{\end{equation}}
\newcommand{\ba}{\begin{array}}
	\newcommand{\ea}{\end{array}}
\newcommand{\bea}{\begin{eqnarray}}
	\newcommand{\eea}{\end{eqnarray}}

\newcommand{\besub}{\begin{subequations}}
	\newcommand{\eesub}{\end{subequations}}

\renewcommand{\eqref}[1]{Eq.~(\ref{eq:#1})}

\def\beq{\begin{equation}}
	\def\eeq#1{\label{#1}\end{equation}}
\def\eeqn{\end{equation}}
\def\beqa{\begin{eqnarray}}
\def\eeqa#1{\label{#1}\end{eqnarray}}
\def\eeqan{\end{eqnarray}}

\definecolor{darkerblue}{rgb}{0.2,0.2,0.5}
\definecolor{seagreen}{rgb}{0.180392,0.545098,0.341176}

\begin{document}

\title{3+1 Dimension Schwinger Pair Production with Quantum Computers}

\author{Bin Xu}
\email{binxu@ufl.edu}  
\author{Wei Xue}
\email{weixue@ufl.edu}  
\affiliation{Department of Physics, University of Florida, Gainesville, FL 32611, USA}



\begin{abstract}

Real-time quantum simulation of quantum field theory in (3+1)D requires large quantum computing resources. 
With a few-qubit quantum computer,
we develop a novel algorithm and experimentally study the Schwinger effect, the electron-positron
pair production in a strong electric field, in (3+1)D.
The resource reduction is achieved by  
treating the electric field as a background field, 
working in Fourier space transverse to the electric field direction, and considering parity symmetry, 
such that we successfully map the three spatial dimension problems into one spatial dimension problems. 
We observe that the rate of pair production of electrons and positrons is consistent with 
the theoretical predication of the Schwinger effect. 
Our work paves the way towards exploring quantum simulation of quantum field theory beyond one spatial dimension.

\end{abstract}
\maketitle

\section{Introduction}
\label{sec:intro}

Historically, the development of quantum electrodynamics~(QED) significantly advances the understanding of quantum 
field theory~(QFT) in its early stage. Quantum simulation of QFT \cite{Jordan:2012xnu, Jordan:2011ci, Jordan:2014tma, Jordan:2017lea, Preskill:2018fag} has a great potential to revolutionize
fundamental physics, in part because it can simulate highly entangled quantum systems, which may never be achieved by 
classical computers.
Currently, it is in its very early stage and QED can play a similar role to make progress in simulating QFT in quantum 
computers.

The Schwinger mechanism \cite{Schwinger:1951nm}, as a textbook example in QED, is employed to improve 
our knowledge of quantum algorithms for QFT. The mechanism describes electron-positron pair production from a 
strong electric field $\bf E$. 
The pair production is viewed as a non-perturbative phenomenon of vacuum decay.
Also, it can be understood from a QED effective action by considering quantum corrections from the interactions of  
the electric field and virtual electron-positron pairs.
It has been invoked to gain insights on
particle production in QCD \cite{Casher:1978wy, Neuberger:1979tb, 
Chiu:1978dg, Glendenning:1983qq, Dosch:1986dp} and on black hole physics
\cite{Hawking:1975vcx, Damour:1976jd}. 
To produce the electron-positron pair in the laboratory, the electric field needs to reach 
the critical value
$E_{critical} \sim 10^{18}\, {\rm volts} / {\rm meter}$, which is extremely strong and not
accessible by current experiments. All these motivate us to study the Schwinger effect 
with quantum computers.

We are in the noisy intermediate-scale quantum (NISQ) era \cite{Preskill2018quantumcomputingin}, in which the 
quantum computing power is limited by its size and the imperfect control of noises. 
Even with the imperfect quantum hardware, some progress of QFT simulation 
has been made in both developing quantum algorithm \cite{Byrnes:2005qx, Wiese:2013uua, Wiese:2014rla, Bermudez:2017yrq, Garcia-Alvarez:2014uda, Zohar:2015hwa, Pichler:2015yqa, Macridin:2018gdw, Klco:2018zqz, Hackett:2018cel, Kreshchuk:2020dla, Haase:2020kaj, Stetina:2020abi, Davoudi:2021ney, Ramirez-Uribe:2021ubp, Stryker:2021asy, Klco:2021lap, Kan:2021xfc} 
and performing quantum simulation \cite{Zohar:2012xf, Zohar:2012ay, Banerjee:2012pg, Banerjee:2012xg, Marcos:2014lda, Zohar:2016iic, Martinez:2016yna, Klco:2018kyo, Bauer:2021gup, Czajka:2021yll, Yeter-Aydeniz:2018mix, Kreshchuk:2020kcz, Li:2021kcs, deJong:2021wsd}.
However, limited by the small-scale quantum computers,
most of the simulations for QFT are demonstrated in (1+1)D.
Going beyond one spatial dimension is a crucial 
step towards exploring more 
fundamental questions in QFT. In this article, we present a novel method to simulate the Schwinger effect in (3+1)D only 
with a few qubits. This method can apply to other field theory questions beyond one spatial dimension.

Because of the significance of the Schwinger mechanism, it is worthwhile to run a real-time simulation
and compare the results to the theoretical predication. 
To simulate the Schwinger pair production, 
we discretize the space along the direction of the electric field and scan the momentum space 
along the direction perpendicular to the electric field. 
Inspired by \cite{Klco:2018kyo}, we impose 
the parity symmetry to further reduce the number of qubits 
by half, and eventually the algorithm is implemented on an IBM's digital quantum computer with 5 qubits. 
The particle-antiparticle pair production is observed in the real-time simulation as expected, and the production rate 
agrees with the predication of the Schwinger effect within uncertainties. 

The article is structured as follows. In \cref{sec:theory} we introduce the theoretical setup for 
quantum simulation of the Schwinger mechanism.
A detailed description of the algorithm is presented in \cref{sec:algorithm}. 
We show the main results of our algorithm and analyze the errors in \cref{sec:result}. 
Finally, we conclude and discuss our anticipation for future works in \cref{sec:conclusion}.

\section{Theoretical Setup}
\label{sec:theory}

In the NISQ era, quantum simulation is limited to a few qubits
and has large quantum noises, which challenges the quantum simulation of QFT in (3+1)D. 
To experimentally study the Schwinger pair production in (3+1)D, we introduce several techniques to simplify and parallelize 
the quantum simulation, including background field method, dimension reduction, parity symmetry, etc. 

We treat the gauge field $A^\mu$ as a classical background field since we are not concerned with the dynamics of $A^\mu$.
A fermion field $\psi$ interacting with the gauge field $A^{\mu}$ is described by the following action,
\begin{equation} 
S= \int {\rm d}^4 x \left[ \bar{\psi} \left( i \slashed{D} -m 
\right)  \psi   
\right]  \ , 
\end{equation} 
where $\slashed{D} = (\partial_\mu + i e A_\mu ) \gamma^\mu$. 
Here we choose the axial gauge with $A_z= 0 $
and $A_0 = - E \, |z|$ to give 
a static electric field ${\bf E}$ in the z-direction.
The background field approach is in accord with the principle of the Euler-Heisenberg Lagrangian 
derived from the effective ation to predict the Schwinger pair creation rate, where $A_\mu$ is a background field and 
fermions are integrated out.
This approach also implies that we neglect the backreaction 
of the produced electrons and positrons to the electric field and interactions among the electrons and positrons.

We observe that the Hamiltonian can be diagonalized by a unitary transformation. Thus a $(3+1)$D simulation is decomposed into 
a summation of several $(1+1)$D simulations, which dramatically 
reduce the number of qubits in the quantum simulation. The reduction is realized by working in the Fourier space of $(x,y)$ and 
the real space of $z$. The Fourier decomposition of fermions takes the form 
\begin{equation}
\psi({\bf x}) =  
\int 
\frac{{\rm d}p_x \, {\rm d}p_y} { ( 2 \pi )^2 }   \,
{\mathsmaller{\sum}}_s
\psi_s (p_x, p_y, z) \, {\rm e}^{i (p_x x  + p_y y ) }  \ , 
\label{eq:phiz}
\end{equation}
with summation over the spin $s$.
By taking a unitary transformation,
\begin{equation}
U = 
\begin{pmatrix}
	\sqrt{ m^\prime + m } &  - \frac{p_x \sigma_x + p_y \sigma_y } {\sqrt{ m^\prime + m }}  \\
	\frac{p_x \sigma_x + p_y \sigma_y } {\sqrt{ m^\prime + m }}  & \sqrt{ m^\prime + m }
\end{pmatrix}   \  
\end{equation}
with $ m^\prime \equiv  \sqrt{ m^2 + p_\perp^2}$ and  
$p_\perp \equiv \sqrt{ p_x^2 + p_y^2} $,
the fermion field is rotated as $\tilde{\psi}_s  (p_x, p_y, z)  =U^\dagger  \psi_s (p_x, p_y, z)   $, 
and the Hamiltonian is converted into a $2$D form 
\begin{eqnarray}
H &=& \int {\rm d} z \frac{{\rm d}p_x \, {\rm d}p_y} { ( 2 \pi )^2 } 
{\mathsmaller{\sum}}_s \, 
\tilde{\psi}_s^\dagger
   \nonumber
   \\
&&
   \left(  \pm i \partial_z \tilde\gamma^0 \tilde\gamma^1 + m' \tilde\gamma^0 + e A_0 \right) 
   \otimes I_{2\times 2} \tilde{\psi}_s \ ,
\end{eqnarray}
where $\pm$ correspond to spin up and down, respectively.
The reduction is achieved by transforming the $4 \times 4$ matrix in the Hamiltonian 
into a $2 \times 2$ $\tilde{\gamma}$ matrix times the identity matrix $I_{2 \times 2}$. 
$\tilde\gamma^i$ happen to have the same form of gamma matrices in $2$D.
In the new Hamiltonian, $ m^\prime $ is an effective mass in $(1+1)$D. Therefore, the Schwinger pair production rate in $(3+1)$D
is given by integrating the rate in $(1+1)$D over the transverse momentum and summing over the spin, 
\begin{equation} 
\Gamma_{3+1} (m ) =  2 \int \frac{{\rm d}^2 p_\perp  } { ( 2 \pi )^2 } 
\Gamma_{1+1} ( m^\prime = \sqrt{ m^2 + p_\perp^2 } ) \ .  
\label{eq:Gamma4d}
\end{equation} 

After reducing to 2D, we spatially discretize the Hamiltonian with the Kogut-Susskind formulation 
\cite{Kogut:1974ag,Banks:1975gq,Casher:1973uf}, mapping electrons (positrons) into even (odd) lattice sites. 
With the lattice spacing $a$, the discretized free Hamiltonian $H_0$ and interaction $H_I$ describing the dynamics of 
one component of fermion field $\phi(n)$ are written as,  
\begin{align}
H_0  & =   H_{0,m} + H_{0,k} = 
m\sum_n(-1)^n\phi^\dagger(n)\phi(n) 
\nonumber\\
& 
+ \frac{i}{2a}\sum_n[\phi^\dagger(n)\phi(n+1)-\phi^\dagger(n+1)\phi(n)] 
\ ,
\\
H_I  & =  \sum_n e A_0(n a)\phi^\dagger(n)\phi(n)   \ . 
\end{align}
where $H_{0,m}$ is the mass term, $H_{0,k}$ is the kinetic term with nearest-neighbor lattice-site interactions.

Furthermore, by considering the parity symmetry of QED, the Hamiltonian is decomposed into the parity even and odd one.
And the parity even/odd fermion fields are $\psi_{\pm} = ( \psi(x) \pm \gamma^0 \psi(-x) )/\sqrt{2} $ for the continuum case 
and the staggered lattice fields $\phi_\pm $ are given in the supplemental material.
The number of spatial sites is reduced by half and the parity even and odd fermions are simulated separately. 
Eventually these parity even or odd fermions are mapped into Pauli spin operators by using the Jordan-Wigner transformation 
\cite{jordan1993paulische}.

\section{Quantum Computing Algorithm}
\label{sec:algorithm}

As shown in the previous section that the $(3+1)$D Schwinger pair production problem is simplified to 1D lattice, 
we proceed to 
simulate this non-perturbative process in digital quantum computers. 
We start from 10 lattice sites $(0, 9)$, where
electron (positron) states are at the even~(odd) lattice sites. 
Then reduce the lattice to $(0, 4)$ by considering parity symmetry. 
The gauge field $A_0 =  n  a   E  $ with $n$ as the lattice number 
$(0,4)$.
Under the parity transformation, $A_0   = ( 10 -n  ) a  E$ in the lattice of $(5,9)$. 

The simulation consists of the following main steps: the ground state preparation, 
time evolution of the parity even Hamiltonian ${\rm e}^{-i H_+ t }$, 
and measurement with the interaction turned off, which is illustrated by the schematic quantum circuit,
\begin{equation}
	\includegraphics[width=0.5\textwidth]{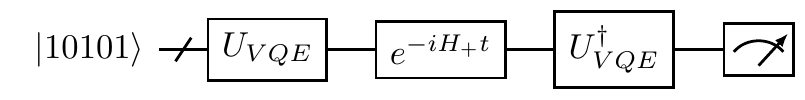}
\nonumber
\end{equation}
Before and after the time evolution, adiabatic turn-on and turn-off of interactions are essential steps, which are
substituted by VQE transformations, $U_{VQE}$ and $U^\dagger_{VQE}$. 
\\

\noindent 
{\bf Ground state:} The ground state of the Hamiltonian with only mass term
$H_{0,m} = \sum_{n=0}^4 (-1)^n m \phi^\dagger (n) \phi(n) $
denoted as $|10101\rangle$ has no electron and positron, which 
is set as the initial state of our simulation.
An excited electron changes $|1\rangle$ to $|0\rangle$ on the 0th, 2nd or 4th qubit, 
while an excited positron changes $|0\rangle$ to $|1\rangle$  on the 1st or 3rd qubit. 
Since charge is conserved, there exists a physical subspace in which the states contain 
three 1's and two 0's given the charge-zero initial state. Such principle turns out to be the 
key to solve the eigenstates of the system and to reduce hardware noises.

After having the initial state $|10101\rangle$, we will turn on the nearest-neighbor lattice-site interactions $H_{0,k}$. 
One approach is to adiabatically turn on the kinetic term $H_{0,k}$ \cite{Jordan:2014tma}, but this is a resource-consuming 
procedure since we need gradually turn on the kinetic term, requiring circuit depth exceeding the limits of hardware
available in the NISQ era. Instead, we use the Variational Quantum Eigensolver (VQE) \cite{peruzzo2014variational} method
to find the ground state of the Hamiltonian $H_0$ with a much shallower circuit. 
The ansatz of VQE circuit mimics the Hamiltonian of the system and contains nine parametrized gates which are rotational operators 
along the $y$ and $z$-axis of the Bloch sphere. 
The 9 parameters are optimized via minimizing the expectation value of the free Hamiltonian $H_0$ to obtain
the vacuum state $| \Omega_{\rm VQE} \rangle$. 
The vacuum state is compared to the exact numerical solutions of the vacuum states of $H_0$, $| \Omega_{\rm exact} \rangle$.
As a result of the VQE method, the fidelity $| \langle    \Omega_{\rm VQE} |  \Omega_{\rm exact} \rangle |^2$ to be larger than $99\%$.  
Hence the VQE method finds a correct vacuum with great accuracy but a few quantum gates.
Note that in the VQE method the operators in the circuit are all charge conserved, such that the states will stay in the 
physical subspace. 
\\

\noindent {\bf Time evolution:} We start with the VQE vacuum state $| \Omega_{\rm VQE} \rangle$ at the beginning of 
the simulation, $t = 0$. The state is evolved with the full Hamiltonian $H = H_0 + H_I$. The parity even(odd) Hamiltonian
$H_+$~($H_-$) is 
a function of Pauli spin operators with the Jordan-Wigner transformation \cite{jordan1993paulische}, 
and is further decomposed into three parts for constructing the quantum algorithm,
\begin{align}
H_{+1}&=\sum_{n=0}^{4}[(-1)^n m+e E a n]\frac{\sigma_3(n)}{2}
\label{eq:H_1}
\\
H_{+2}&=\frac{1}{\sqrt{2}a}[\sigma^+(0)\sigma^-(1)+\sigma^+(1)\sigma^-(0)] 
\nonumber \\
&  +\frac{1}{2a}[\sigma^+(2)\sigma^-(3)+\sigma^+(3)\sigma^-(2)]
\label{eq:H_2}
\\
H_{+3}&=\frac{1}{2a}\sum_{n=1,3}[\sigma^+(n)\sigma^-(n+1)+\sigma^+(n+1)\sigma^-(n)] \ , 
\label{eq:Hdecomposition}
\end{align}
with $\sigma^{\pm}(n)=[\sigma_1(n)\pm i\sigma_2(n)]/2$. 
The unitary time evolution $U(t)$ can be expanded by using 
Suzuki-Trotter formulae with $n_t$ steps \cite{trotter1959product, Suzuki:1976be},
\begin{equation}
U(t) = {\rm e}^{-i H_+ t}=\left(\prod_{j=1}^3 e^{-i H_{+j} \delta t}\right)^{n_t}+ {\cal O}(  t^2 /  n_t) \ , 
\end{equation}
where $\delta t = t / n_t$.
The Hamiltonian decomposition in \cref{eq:H_1,eq:H_2,eq:Hdecomposition} is optimized by the quantum gate implementation.
For each time step, $\exp( - i H_{+1} \delta t )$ is implemented by the rotation gate along z-axis, $R_z$, 
while $\exp( - i H_{+2} \delta t )$ and $\exp( - i H_{+3} \delta t )$ are realized by two CNOT and one controlled x-axis rotation gate $R_x$.   
In the simulation, number of time steps is determined by trading off between Trotterization errors and hardware quantum noises.
\\

\noindent{\bf Measurement:}
After $n_t$ steps of evolution, we should turn off the kinetic term adiabatically then measure the probability of being the vacuum state. 
Such adiabatic turn-off is equivalent to applying an inverse VQE operator that we employed previously to make $|\Omega_{VQE}\rangle$.
The vacuum persistence probability is measured by the frequency of the state $|10101\rangle$.

Although all of the previous procedures in our algorithm are charge conserved, 
the quantum noises can give non-zero charge final states.
We propose to consider the physical states and remove the non-zero charge states because it will reduce the hardware errors.
Suppose that the probability of a single bit flip error is of $O(\epsilon)$, then 
it needs at least two bit flips to return to the physical subspace. Therefore, when we measure the vacuum persistence probability by 
the frequency of $|10101\rangle$ restricted in the physical subspace, the hardware error is suppressed from 
$O(\epsilon)$ to $O(\epsilon^2)$.

\section{Schwinger pair production results}
\label{sec:result}

In this section, we analyze the numerical results of the Schwinger pair production with IBM quantum computers.
Our analysis is based on the simulations running on an IBM machine, {\it ibm\_lagos}, 
and on the simulations given by simulators from QISKIT \cite{abraham2019qiskit} by turning off the quantum noises.
The quantum simulation results 
are compared to the theoretical predication, and 
they are consistent within the experimental uncertainties.
On the IBM quantum computer, following the dimension reduction method in the previous sections, we simulate the 
(1+1)D Schwinger effect with different effective masses $m'=\sqrt{m^2+p_\perp^2}$ in parallel. 
By repeating the experiments and measuring the final states a large number of times,
we deduce the vacuum persistent probability $P_{vac}(t)$ and the (1+1)D vacuum decay rate $\Gamma_{1+1}$. 
The decay rate in (3+1)D is obtained by integrating the decay rate $\Gamma_{1+1}$ over the transverse momentum as \cref{eq:Gamma4d}.
In the end, we discuss the uncertainties from the quantum simulation.

A benchmark point is chosen to demonstrate the validity of the quantum algorithm, 
where the parameters are given as 
\begin{equation}
e E=20 \, , \quad
	a=0.45 \, , 
\end{equation}
the number of lattice sites (qubits) $N=5$ and the time step $n_t$ = 3 for the Suzuki-Trotter expansion. 
Here we set $m= 1$ as a unit, the spacing $a $ is determined by restricting the error within $10\%$, 
and $eE = 20 $ represents a strong electric field. A weaker electric field takes a longer time in simulations to 
observe the vacuum decay, while a much stronger electric field will quickly populate quite a few electrons and positrons,
such that the correction to the decay rate is not negligible both in theory and experiments. Hence we choose this intermediate
value of $eE$.

\begin{figure}[!h]
	\centering
	\includegraphics[width=0.5\textwidth]{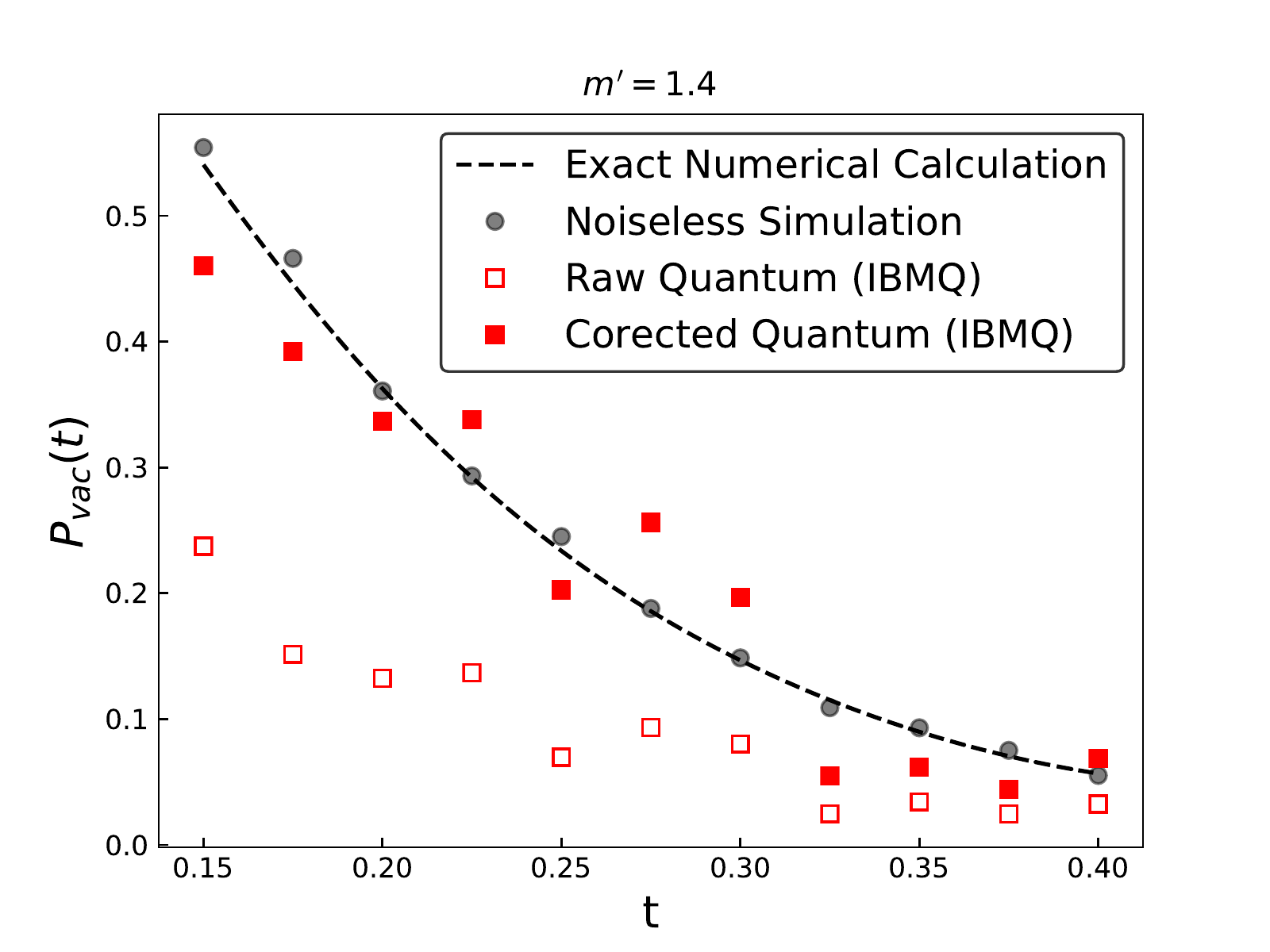}
	\caption{Vacuum persistent probability for the effective mass $m' = 1.4$.
The dashed curve corresponds to the exact numerical solution of the Hamiltonian. 
The simulation results on a noiseless simulator are shown in the gray dots. 
The {\it ibm\_lagos} quantum computer results are shown in the hallow square.  
The solid squares are the corrected results to the quantum simulation by restricting to the charger-conserving subspace.}
	\label{fig1+1d}
\end{figure}

The probability of the final states being the vacuum state $|10101\rangle$
denoted as the vacuum persistence probability $P_{vac}(t)$, is measured at different times. 
As an example, the probability for $m' = 1.4$ is shown in \cref{fig1+1d}, and $P_{vac}(t)$ for other masses is
presented in the supplemental material.
In \cref{fig1+1d}, the numerical solution of the Hamiltonian in \cref{eq:H_1,eq:H_2,eq:Hdecomposition}
is shown in the dotted line.
The numerical calculation gives the expected results for simulations to compare to. 
Also, it tells whether the parameters, $a$, $eE$ and $N$ is a proper choice by comparing 
to the Schwinger pair production rate derived from 
the continuous spacetime, which is given as \cite{Cohen:2008wz}  
\begin{equation}
	\Gamma_{1+1}(m)=\frac{e E}{2\pi}\log(1-e^{-\frac{\pi m^2}{e E}}) \ .
   \label{eq:GammaM}
\end{equation}
We perform the simulation in two systems: IBM's quantum simulators and quantum computers.
The IBM's simulators are designed to simulate and test the quantum computers,
where we turn off quantum noises in the simulators to validate our quantum algorithm in an ideal situation. 
As shown in \cref{fig1+1d}, the noiseless
simulation results are aligned with the ``exact numerical calculation''.
The quantum simulation results shown in the red hollow squares deviate from the exact solution due to the large 
noises. After removing the states having non-zero charges to restrict the final states in the physical subspace, 
we obtain the corrected quantum simulation results close to the theoretical ones.

\begin{figure}[!h]
	\centering
	\includegraphics[width=0.5\textwidth]{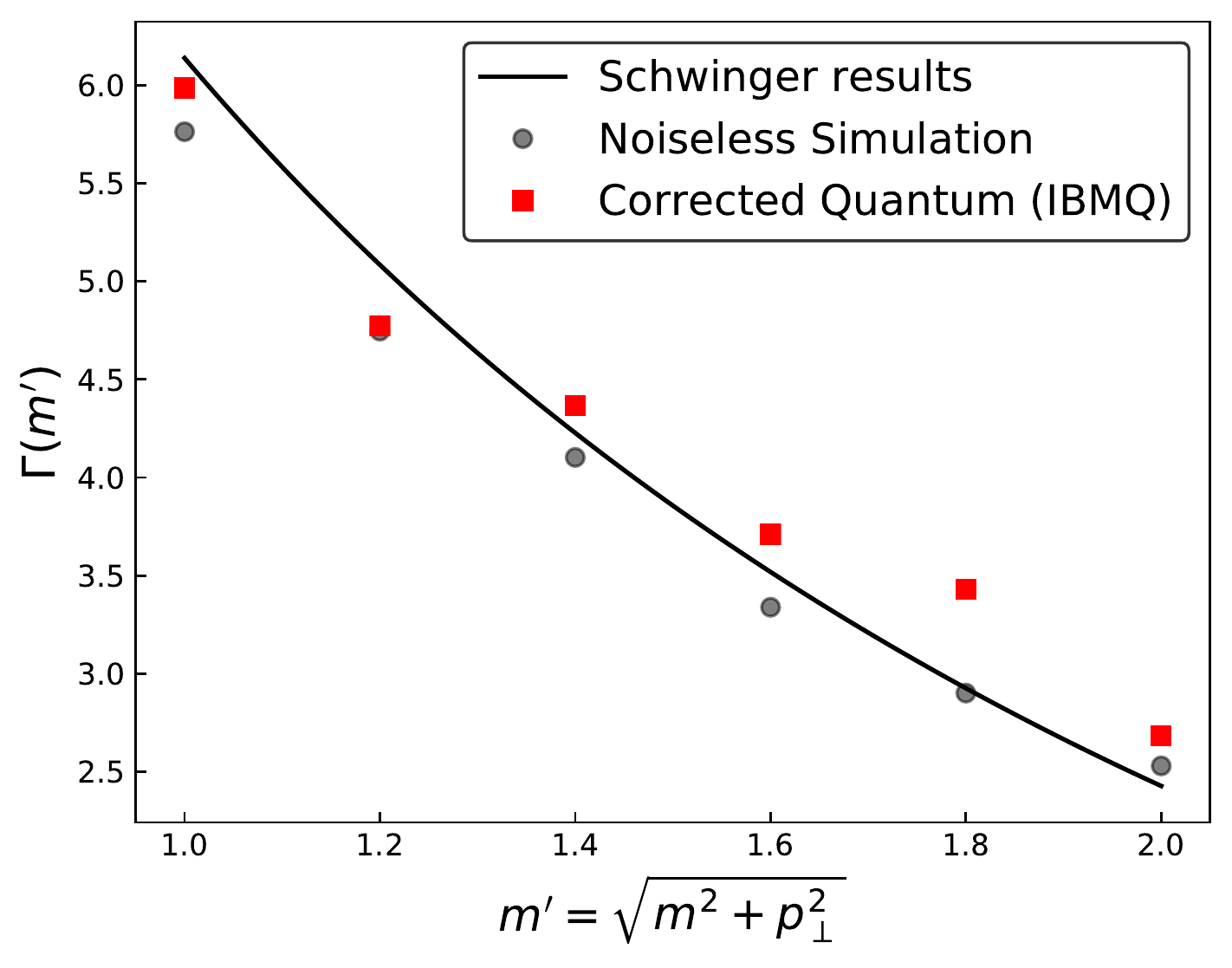}
	\caption{Vacuum decay rates versus effective masses $m'$. 
The solid curve corresponds to the theoretical predications from QED in continuous space-time. 
The fitting results of the noiseless simulator are shown in the gray dots.
The results from the IBM {\it ibm\_lagos} are shown in the square data points. }
\label{fig3+1d}
\end{figure}

For a given effective mass $m'$, the persistent vacuum probability $P_{vac}(t)$ data are fitted by an exponential
function of $ c_1 \exp ( - \Gamma_{1+1}(m') \times  volume \times t )$ with a normalization factor $c_1$ and 
$volume = a N$.
The fitting results give the decay rate per volume $\Gamma_{1+1}(m')$ and are shown in \cref{fig3+1d}. 
Note that we choose a time range to fit the exponential curve and the normalization $c_1$ is set as a free parameter.
In the case of $m' =1.4$, the time region is taken as $0.10<t<0.45$; for other masses, the range is given in the supplemental 
material.
The time range is set by considering that
there is a transient effect at the beginning of time due to suddenly turning on the electric field
and back reactions for a large time due to finite spatial size. 

Integrating the $\Gamma_{1+1}(m')$ over the transverse momentum as \cref{eq:Gamma4d} gives the electron-positron pair 
production rate in (3+1)D. 
Here we only consider the electron-positron pair production with the transverse momentum $p_\perp^2 \leq 3$, i.e. $m' \leq 2$, 
due to larger quantum noises for larger $m'$.
The decay rate from the theoretical predication of QED is $0.58$, the noiseless simulator result is $0.56$, and 
the corrected quantum computer gives the decay rate $0.60$.

There are various uncertainties in the quantum simulation, originated from space and time discretization, statistical error,
and hardware noises, in which the hardware noises are the dominant ones. 
First, modeling the Schwinger effect in a lattice as \cref{eq:H_1,eq:H_2,eq:Hdecomposition} introduces discretization error and 
truncation error by choosing the spacing $a$ and the finite size $L=N a$. The errors from spacing and finite size 
are of orders ${\cal O} (a)$ and ${\cal O}(1/L^2)$, respectively.
Here taking $N=5, a=0.45$ in the benchmark model gives the numerical solution of the Hamiltonian close to the theoretical predication
of QED, which can be seen from comparing the Schwinger results to the noiseless simulation results in \cref{fig3+1d}.
Second, we experience statistical error of order ${\cal O}(1/\sqrt{n_{shot}})$ either on a simulator or a quantum computer with 
$n_{shot}$ measurement shots. We take $n_{shot}=8192$ to make sure the statistical error is subdominant compared to the other ones.
Third,
$n_t$ Trotter steps in time
introduces an error of order ${\cal O}(t^2/n_t)$. Here we choose $n_t=3$ to give accurate enough simulation results and moderate 
circuit depth. 
Finally, hardware noises, especially readout errors and CNOT gate errors, in the NISQ era are quite large. 
On the {\it ibm\_lagos} quantum computer the readout errors and CNOT gate errors are both about $1\%$. 
The VQE transformation as well as each Trotter step takes 8 CNOT gates, such that the whole circuit contains 40 CNOT gates.  
Hardware noises are relieved by restricting to the charge-conserved subspace. 
We could also apply readout error mitigation\cite{nachman2020unfolding} or CNOT gate noise mitigation\cite{he2020resource} to further reduce 
hardware noise, which we would like to explore in future works.

\section{Conclusion}
\label{sec:conclusion}

In this article, we demonstrate a quantum algorithm to simulate the non-perturbative phenomenon of the Schwinger effect in (3+1)D 
by using IBM's digital quantum computers. The number of qubits, the depth of the circuit are reduced suitable to the NISQ area, 
and the quantum noise is under control to get the Schwinger pair production rate.
We manage to perform the quantum simulation in 5-qubit, by treating the gauge field $A^\mu$
as a background field, finding the Hamiltonian diagonalization and further reducing the resources by parity symmetry.
In the real-time quantum simulation, we prepare the ground state, evolve it to a given time with a few time-step and 
measure the final states. 
The depth of the circuit is shortened by implementing the VQE algorithm instead of adiabatic turn on(off).
In the measurement, the hardware noises are relieved by restricting to the charge-conserving final states. 
We implement the algorithm on IBM's quantum computers also in the noiseless simulators. By analyzing the results and comparing them
to the exact numerical calculation and theoretical predication of the Schwinger effect, we conclude that the results are consistent
within uncertainties. This methodology can apply to some quantum field theory questions related to effective actions. 
We look forward to future studies of quantum simulation of QFT beyond one spatial dimension.

\begin{acknowledgments}
We are grateful to Pierre Ramond
for useful discussions.
We acknowledge use of the IBM Q experience for this work. 
B.X and W.X. are supported in part by the DOE grant DE-SC0010296.
\end{acknowledgments} 

\bibliographystyle{utphys}
\bibliography{ref}

\providecommand{\href}[2]{#2}\begingroup\raggedright\begin{thebibliography}{10}

\bibitem{Jordan:2012xnu}
S.~P. Jordan, K.~S.~M. Lee, and J.~Preskill, ``{Quantum Algorithms for Quantum
  Field Theories},'' \href{http://dx.doi.org/10.1126/science.1217069}{{\em
  Science} {\bfseries 336} (2012) 1130--1133},
  \href{http://arxiv.org/abs/1111.3633}{{\ttfamily arXiv:1111.3633
  [quant-ph]}}.

\bibitem{Jordan:2011ci}
S.~P. Jordan, K.~S.~M. Lee, and J.~Preskill, ``{Quantum Computation of
  Scattering in Scalar Quantum Field Theories},'' {\em Quant. Inf. Comput.}
  {\bfseries 14} (2014) 1014--1080,
  \href{http://arxiv.org/abs/1112.4833}{{\ttfamily arXiv:1112.4833 [hep-th]}}.

\bibitem{Jordan:2014tma}
S.~P. Jordan, K.~S.~M. Lee, and J.~Preskill, ``{Quantum Algorithms for
  Fermionic Quantum Field Theories},''
  \href{http://arxiv.org/abs/1404.7115}{{\ttfamily arXiv:1404.7115 [hep-th]}}.

\bibitem{Jordan:2017lea}
S.~P. Jordan, H.~Krovi, K.~S.~M. Lee, and J.~Preskill, ``{BQP-completeness of
  Scattering in Scalar Quantum Field Theory},''
  \href{http://dx.doi.org/10.22331/q-2018-01-08-44}{{\em Quantum} {\bfseries 2}
  (2018) 44}, \href{http://arxiv.org/abs/1703.00454}{{\ttfamily
  arXiv:1703.00454 [quant-ph]}}.

\bibitem{Preskill:2018fag}
J.~Preskill, ``{Simulating quantum field theory with a quantum computer},''
  \href{http://dx.doi.org/10.22323/1.334.0024}{{\em PoS} {\bfseries
  LATTICE2018} (2018) 024}, \href{http://arxiv.org/abs/1811.10085}{{\ttfamily
  arXiv:1811.10085 [hep-lat]}}.

\bibitem{Schwinger:1951nm}
J.~S. Schwinger, ``{On gauge invariance and vacuum polarization},''
  \href{http://dx.doi.org/10.1103/PhysRev.82.664}{{\em Phys. Rev.} {\bfseries
  82} (1951) 664--679}.

\bibitem{Casher:1978wy}
A.~Casher, H.~Neuberger, and S.~Nussinov, ``{Chromoelectric Flux Tube Model of
  Particle Production},'' \href{http://dx.doi.org/10.1103/PhysRevD.20.179}{{\em
  Phys. Rev. D} {\bfseries 20} (1979) 179--188}.

\bibitem{Neuberger:1979tb}
H.~Neuberger, ``{Finite Time Corrections to the Chromoelectric Flux Tube
  Model},'' \href{http://dx.doi.org/10.1103/PhysRevD.20.2936}{{\em Phys. Rev.
  D} {\bfseries 20} (1979) 2936}.

\bibitem{Chiu:1978dg}
C.~B. Chiu and S.~Nussinov, ``{A Diagrammatic Approach to Pair Production in
  Slowly Varying and Constant Fields},''
  \href{http://dx.doi.org/10.1103/PhysRevD.20.945}{{\em Phys. Rev. D}
  {\bfseries 20} (1979) 945}.

\bibitem{Glendenning:1983qq}
N.~K. Glendenning and T.~Matsui, ``{CREATION OF ANTI-Q Q PAIR IN A
  CHROMOELECTRIC FLUX TUBE},''
  \href{http://dx.doi.org/10.1103/PhysRevD.28.2890}{{\em Phys. Rev. D}
  {\bfseries 28} (1983) 2890--2891}.

\bibitem{Dosch:1986dp}
H.~G. Dosch and D.~Gromes, ``{Theoretical Foundation for Treating Decays
  Allowed by the Okubo-zweig-iizuka Rule and Related Phenomena},''
  \href{http://dx.doi.org/10.1103/PhysRevD.33.1378}{{\em Phys. Rev. D}
  {\bfseries 33} (1986) 1378--1386}.

\bibitem{Hawking:1975vcx}
S.~W. Hawking, ``{Particle Creation by Black Holes},''
  \href{http://dx.doi.org/10.1007/BF02345020}{{\em Commun. Math. Phys.}
  {\bfseries 43} (1975) 199--220}. [Erratum: Commun.Math.Phys. 46, 206 (1976)].

\bibitem{Damour:1976jd}
T.~Damour and R.~Ruffini, ``{Black Hole Evaporation in the
  Klein-Sauter-Heisenberg-Euler Formalism},''
  \href{http://dx.doi.org/10.1103/PhysRevD.14.332}{{\em Phys. Rev. D}
  {\bfseries 14} (1976) 332--334}.

\bibitem{Preskill2018quantumcomputingin}
J.~Preskill, ``Quantum {C}omputing in the {NISQ} era and beyond,''
  \href{http://dx.doi.org/10.22331/q-2018-08-06-79}{{\em {Quantum}} {\bfseries
  2} (Aug., 2018) 79}. \url{https://doi.org/10.22331/q-2018-08-06-79}.

\bibitem{Byrnes:2005qx}
T.~Byrnes and Y.~Yamamoto, ``{Simulating lattice gauge theories on a quantum
  computer},'' \href{http://dx.doi.org/10.1103/PhysRevA.73.022328}{{\em Phys.
  Rev. A} {\bfseries 73} (2006) 022328},
  \href{http://arxiv.org/abs/quant-ph/0510027}{{\ttfamily
  arXiv:quant-ph/0510027}}.

\bibitem{Wiese:2013uua}
U.-J. Wiese, ``{Ultracold Quantum Gases and Lattice Systems: Quantum Simulation
  of Lattice Gauge Theories},''
  \href{http://dx.doi.org/10.1002/andp.201300104}{{\em Annalen Phys.}
  {\bfseries 525} (2013) 777--796},
  \href{http://arxiv.org/abs/1305.1602}{{\ttfamily arXiv:1305.1602
  [quant-ph]}}.

\bibitem{Wiese:2014rla}
U.-J. Wiese, ``{Towards Quantum Simulating QCD},''
  \href{http://dx.doi.org/10.1016/j.nuclphysa.2014.09.102}{{\em Nucl. Phys. A}
  {\bfseries 931} (2014) 246--256},
  \href{http://arxiv.org/abs/1409.7414}{{\ttfamily arXiv:1409.7414 [hep-th]}}.

\bibitem{Bermudez:2017yrq}
A.~Bermudez, G.~Aarts, and M.~M\"uller, ``{Quantum sensors for the generating
  functional of interacting quantum field theories},''
  \href{http://dx.doi.org/10.1103/PhysRevX.7.041012}{{\em Phys. Rev. X}
  {\bfseries 7} no.~4, (2017) 041012},
  \href{http://arxiv.org/abs/1704.02877}{{\ttfamily arXiv:1704.02877
  [quant-ph]}}.

\bibitem{Garcia-Alvarez:2014uda}
L.~Garc\'\i{}a-\'Alvarez, J.~Casanova, A.~Mezzacapo, I.~L. Egusquiza,
  L.~Lamata, G.~Romero, and E.~Solano, ``{Fermion-Fermion Scattering in Quantum
  Field Theory with Superconducting Circuits},''
  \href{http://dx.doi.org/10.1103/PhysRevLett.114.070502}{{\em Phys. Rev.
  Lett.} {\bfseries 114} no.~7, (2015) 070502},
  \href{http://arxiv.org/abs/1404.2868}{{\ttfamily arXiv:1404.2868
  [quant-ph]}}.

\bibitem{Zohar:2015hwa}
E.~Zohar, J.~I. Cirac, and B.~Reznik, ``{Quantum Simulations of Lattice Gauge
  Theories using Ultracold Atoms in Optical Lattices},''
  \href{http://dx.doi.org/10.1088/0034-4885/79/1/014401}{{\em Rept. Prog.
  Phys.} {\bfseries 79} no.~1, (2016) 014401},
  \href{http://arxiv.org/abs/1503.02312}{{\ttfamily arXiv:1503.02312
  [quant-ph]}}.

\bibitem{Pichler:2015yqa}
T.~Pichler, M.~Dalmonte, E.~Rico, P.~Zoller, and S.~Montangero, ``{Real-time
  Dynamics in U(1) Lattice Gauge Theories with Tensor Networks},''
  \href{http://dx.doi.org/10.1103/PhysRevX.6.011023}{{\em Phys. Rev. X}
  {\bfseries 6} no.~1, (2016) 011023},
  \href{http://arxiv.org/abs/1505.04440}{{\ttfamily arXiv:1505.04440
  [cond-mat.quant-gas]}}.

\bibitem{Macridin:2018gdw}
A.~Macridin, P.~Spentzouris, J.~Amundson, and R.~Harnik, ``{Electron-Phonon
  Systems on a Universal Quantum Computer},''
  \href{http://dx.doi.org/10.1103/PhysRevLett.121.110504}{{\em Phys. Rev.
  Lett.} {\bfseries 121} no.~11, (2018) 110504},
  \href{http://arxiv.org/abs/1802.07347}{{\ttfamily arXiv:1802.07347
  [quant-ph]}}.

\bibitem{Klco:2018zqz}
N.~Klco and M.~J. Savage, ``{Digitization of scalar fields for quantum
  computing},'' \href{http://dx.doi.org/10.1103/PhysRevA.99.052335}{{\em Phys.
  Rev. A} {\bfseries 99} no.~5, (2019) 052335},
  \href{http://arxiv.org/abs/1808.10378}{{\ttfamily arXiv:1808.10378
  [quant-ph]}}.

\bibitem{Hackett:2018cel}
D.~C. Hackett, K.~Howe, C.~Hughes, W.~Jay, E.~T. Neil, and J.~N. Simone,
  ``{Digitizing Gauge Fields: Lattice Monte Carlo Results for Future Quantum
  Computers},'' \href{http://dx.doi.org/10.1103/PhysRevA.99.062341}{{\em Phys.
  Rev. A} {\bfseries 99} no.~6, (2019) 062341},
  \href{http://arxiv.org/abs/1811.03629}{{\ttfamily arXiv:1811.03629
  [quant-ph]}}.

\bibitem{Kreshchuk:2020dla}
M.~Kreshchuk, W.~M. Kirby, G.~Goldstein, H.~Beauchemin, and P.~J. Love,
  ``{Quantum Simulation of Quantum Field Theory in the Light-Front
  Formulation},'' \href{http://arxiv.org/abs/2002.04016}{{\ttfamily
  arXiv:2002.04016 [quant-ph]}}.

\bibitem{Haase:2020kaj}
J.~F. Haase, L.~Dellantonio, A.~Celi, D.~Paulson, A.~Kan, K.~Jansen, and C.~A.
  Muschik, ``{A resource efficient approach for quantum and classical
  simulations of gauge theories in particle physics},''
  \href{http://dx.doi.org/10.22331/q-2021-02-04-393}{{\em Quantum} {\bfseries
  5} (2021) 393}, \href{http://arxiv.org/abs/2006.14160}{{\ttfamily
  arXiv:2006.14160 [quant-ph]}}.

\bibitem{Stetina:2020abi}
T.~F. Stetina, A.~Ciavarella, X.~Li, and N.~Wiebe, ``{Simulating Effective QED
  on Quantum Computers},'' \href{http://arxiv.org/abs/2101.00111}{{\ttfamily
  arXiv:2101.00111 [quant-ph]}}.

\bibitem{Davoudi:2021ney}
Z.~Davoudi, N.~M. Linke, and G.~Pagano, ``{Toward simulating quantum field
  theories with controlled phonon-ion dynamics: A hybrid analog-digital
  approach},'' \href{http://dx.doi.org/10.1103/PhysRevResearch.3.043072}{{\em
  Phys. Rev. Res.} {\bfseries 3} no.~4, (2021) 043072},
  \href{http://arxiv.org/abs/2104.09346}{{\ttfamily arXiv:2104.09346
  [quant-ph]}}.

\bibitem{Ramirez-Uribe:2021ubp}
S.~Ram\'\i{}rez-Uribe, A.~E. Renter\'\i{}a-Olivo, G.~Rodrigo, G.~F.~R.
  Sborlini, and L.~Vale~Silva, ``{Quantum algorithm for Feynman loop
  integrals},'' \href{http://arxiv.org/abs/2105.08703}{{\ttfamily
  arXiv:2105.08703 [hep-ph]}}.

\bibitem{Stryker:2021asy}
J.~R. Stryker, ``{Shearing approach to gauge invariant Trotterization},''
  \href{http://arxiv.org/abs/2105.11548}{{\ttfamily arXiv:2105.11548
  [hep-lat]}}.

\bibitem{Klco:2021lap}
N.~Klco, A.~Roggero, and M.~J. Savage, ``{Standard Model Physics and the
  Digital Quantum Revolution: Thoughts about the Interface},''
  \href{http://arxiv.org/abs/2107.04769}{{\ttfamily arXiv:2107.04769
  [quant-ph]}}.

\bibitem{Kan:2021xfc}
A.~Kan and Y.~Nam, ``{Lattice Quantum Chromodynamics and Electrodynamics on a
  Universal Quantum Computer},''
  \href{http://arxiv.org/abs/2107.12769}{{\ttfamily arXiv:2107.12769
  [quant-ph]}}.

\bibitem{Zohar:2012xf}
E.~Zohar, J.~I. Cirac, and B.~Reznik, ``{Cold-Atom Quantum Simulator for SU(2)
  Yang-Mills Lattice Gauge Theory},''
  \href{http://dx.doi.org/10.1103/PhysRevLett.110.125304}{{\em Phys. Rev.
  Lett.} {\bfseries 110} no.~12, (2013) 125304},
  \href{http://arxiv.org/abs/1211.2241}{{\ttfamily arXiv:1211.2241
  [quant-ph]}}.

\bibitem{Zohar:2012ay}
E.~Zohar, J.~I. Cirac, and B.~Reznik, ``{Simulating Compact Quantum
  Electrodynamics with ultracold atoms: Probing confinement and nonperturbative
  effects},'' \href{http://dx.doi.org/10.1103/PhysRevLett.109.125302}{{\em
  Phys. Rev. Lett.} {\bfseries 109} (2012) 125302},
  \href{http://arxiv.org/abs/1204.6574}{{\ttfamily arXiv:1204.6574
  [quant-ph]}}.

\bibitem{Banerjee:2012pg}
D.~Banerjee, M.~Dalmonte, M.~Muller, E.~Rico, P.~Stebler, U.~J. Wiese, and
  P.~Zoller, ``{Atomic Quantum Simulation of Dynamical Gauge Fields coupled to
  Fermionic Matter: From String Breaking to Evolution after a Quench},''
  \href{http://dx.doi.org/10.1103/PhysRevLett.109.175302}{{\em Phys. Rev.
  Lett.} {\bfseries 109} (2012) 175302},
  \href{http://arxiv.org/abs/1205.6366}{{\ttfamily arXiv:1205.6366
  [cond-mat.quant-gas]}}.

\bibitem{Banerjee:2012xg}
D.~Banerjee, M.~B\"ogli, M.~Dalmonte, E.~Rico, P.~Stebler, U.~J. Wiese, and
  P.~Zoller, ``{Atomic Quantum Simulation of U(N) and SU(N) Non-Abelian Lattice
  Gauge Theories},''
  \href{http://dx.doi.org/10.1103/PhysRevLett.110.125303}{{\em Phys. Rev.
  Lett.} {\bfseries 110} no.~12, (2013) 125303},
  \href{http://arxiv.org/abs/1211.2242}{{\ttfamily arXiv:1211.2242
  [cond-mat.quant-gas]}}.

\bibitem{Marcos:2014lda}
D.~Marcos, P.~Widmer, E.~Rico, M.~Hafezi, P.~Rabl, U.~J. Wiese, and P.~Zoller,
  ``{Two-dimensional Lattice Gauge Theories with Superconducting Quantum
  Circuits},'' \href{http://dx.doi.org/10.1016/j.aop.2014.09.011}{{\em Annals
  Phys.} {\bfseries 351} (2014) 634--654},
  \href{http://arxiv.org/abs/1407.6066}{{\ttfamily arXiv:1407.6066
  [quant-ph]}}.

\bibitem{Zohar:2016iic}
E.~Zohar, A.~Farace, B.~Reznik, and J.~I. Cirac, ``{Digital lattice gauge
  theories},'' \href{http://dx.doi.org/10.1103/PhysRevA.95.023604}{{\em Phys.
  Rev. A} {\bfseries 95} no.~2, (2017) 023604},
  \href{http://arxiv.org/abs/1607.08121}{{\ttfamily arXiv:1607.08121
  [quant-ph]}}.

\bibitem{Martinez:2016yna}
E.~A. Martinez {\em et~al.}, ``{Real-time dynamics of lattice gauge theories
  with a few-qubit quantum computer},''
  \href{http://dx.doi.org/10.1038/nature18318}{{\em Nature} {\bfseries 534}
  (2016) 516--519}, \href{http://arxiv.org/abs/1605.04570}{{\ttfamily
  arXiv:1605.04570 [quant-ph]}}.

\bibitem{Klco:2018kyo}
N.~Klco, E.~F. Dumitrescu, A.~J. McCaskey, T.~D. Morris, R.~C. Pooser, M.~Sanz,
  E.~Solano, P.~Lougovski, and M.~J. Savage, ``{Quantum-classical computation
  of Schwinger model dynamics using quantum computers},''
  \href{http://dx.doi.org/10.1103/PhysRevA.98.032331}{{\em Phys. Rev. A}
  {\bfseries 98} no.~3, (2018) 032331},
  \href{http://arxiv.org/abs/1803.03326}{{\ttfamily arXiv:1803.03326
  [quant-ph]}}.

\bibitem{Bauer:2021gup}
C.~W. Bauer, M.~Freytsis, and B.~Nachman, ``{Simulating Collider Physics on
  Quantum Computers Using Effective Field Theories},''
  \href{http://dx.doi.org/10.1103/PhysRevLett.127.212001}{{\em Phys. Rev.
  Lett.} {\bfseries 127} no.~21, (2021) 212001},
  \href{http://arxiv.org/abs/2102.05044}{{\ttfamily arXiv:2102.05044
  [hep-ph]}}.

\bibitem{Czajka:2021yll}
A.~M. Czajka, Z.-B. Kang, H.~Ma, and F.~Zhao, ``{Quantum Simulation of Chiral
  Phase Transitions},'' \href{http://arxiv.org/abs/2112.03944}{{\ttfamily
  arXiv:2112.03944 [hep-ph]}}.

\bibitem{Yeter-Aydeniz:2018mix}
K.~Yeter-Aydeniz, E.~F. Dumitrescu, A.~J. McCaskey, R.~S. Bennink, R.~C.
  Pooser, and G.~Siopsis, ``{Scalar Quantum Field Theories as a Benchmark for
  Near-Term Quantum Computers},''
  \href{http://dx.doi.org/10.1103/PhysRevA.99.032306}{{\em Phys. Rev. A}
  {\bfseries 99} no.~3, (2019) 032306},
  \href{http://arxiv.org/abs/1811.12332}{{\ttfamily arXiv:1811.12332
  [quant-ph]}}.

\bibitem{Kreshchuk:2020kcz}
M.~Kreshchuk, S.~Jia, W.~M. Kirby, G.~Goldstein, J.~P. Vary, and P.~J. Love,
  ``{Light-Front Field Theory on Current Quantum Computers},''
  \href{http://dx.doi.org/10.3390/e23050597}{{\em Entropy} {\bfseries 23}
  no.~5, (2021) 597}, \href{http://arxiv.org/abs/2009.07885}{{\ttfamily
  arXiv:2009.07885 [quant-ph]}}.

\bibitem{Li:2021kcs}
T.~Li, X.~Guo, W.~K. Lai, X.~Liu, E.~Wang, H.~Xing, D.-B. Zhang, and S.-L. Zhu,
  ``{Partonic Structure by Quantum Computing},''
  \href{http://arxiv.org/abs/2106.03865}{{\ttfamily arXiv:2106.03865
  [hep-ph]}}.

\bibitem{deJong:2021wsd}
W.~A. de~Jong, K.~Lee, J.~Mulligan, M.~P\l{}osko\'n, F.~Ringer, and X.~Yao,
  ``{Quantum simulation of non-equilibrium dynamics and thermalization in the
  Schwinger model},'' \href{http://arxiv.org/abs/2106.08394}{{\ttfamily
  arXiv:2106.08394 [quant-ph]}}.

\bibitem{Kogut:1974ag}
J.~B. Kogut and L.~Susskind, ``{Hamiltonian Formulation of Wilson's Lattice
  Gauge Theories},'' \href{http://dx.doi.org/10.1103/PhysRevD.11.395}{{\em
  Phys. Rev. D} {\bfseries 11} (1975) 395--408}.

\bibitem{Banks:1975gq}
T.~Banks, L.~Susskind, and J.~B. Kogut, ``{Strong Coupling Calculations of
  Lattice Gauge Theories: (1+1)-Dimensional Exercises},''
  \href{http://dx.doi.org/10.1103/PhysRevD.13.1043}{{\em Phys. Rev. D}
  {\bfseries 13} (1976) 1043}.

\bibitem{Casher:1973uf}
A.~Casher, J.~B. Kogut, and L.~Susskind, ``{Vacuum polarization and the quark
  parton puzzle},'' \href{http://dx.doi.org/10.1103/PhysRevLett.31.792}{{\em
  Phys. Rev. Lett.} {\bfseries 31} (1973) 792--795}.

\bibitem{jordan1993paulische}
P.~Jordan and E.~P. Wigner, ``{\"u}ber das paulische {\"a}quivalenzverbot,'' in
  {\em The Collected Works of Eugene Paul Wigner}, pp.~109--129.
\newblock Springer, 1993.

\bibitem{peruzzo2014variational}
A.~Peruzzo, J.~McClean, P.~Shadbolt, M.-H. Yung, X.-Q. Zhou, P.~J. Love,
  A.~Aspuru-Guzik, and J.~L. O’brien, ``A variational eigenvalue solver on a
  photonic quantum processor,''
  \href{http://dx.doi.org/10.1038/ncomms5213}{{\em Nature communications}
  {\bfseries 5} no.~1, (2014) 1--7}.

\bibitem{trotter1959product}
H.~F. Trotter, ``On the product of semi-groups of operators,''
  \href{http://dx.doi.org/10.2307/2033649}{{\em Proceedings of the American
  Mathematical Society} {\bfseries 10} no.~4, (1959) 545--551}.

\bibitem{Suzuki:1976be}
M.~Suzuki, ``{Generalized Trotter's Formula and Systematic Approximants of
  Exponential Operators and Inner Derivations with Applications to Many Body
  Problems},'' \href{http://dx.doi.org/10.1007/BF01609348}{{\em Commun. Math.
  Phys.} {\bfseries 51} (1976) 183--190}.

\bibitem{abraham2019qiskit}
H.~Abraham, I.~Y. Akhalwaya, G.~Aleksandrowicz, T.~Alexander, G.~Alexandrowics,
  E.~Arbel, A.~Asfaw, C.~Azaustre, P.~Barkoutsos, G.~Barron, {\em et~al.},
  ``Qiskit: An open-source framework for quantum computing, 2019,'' {\em URL
  https://doi. org/10.5281/zenodo} {\bfseries 2562111} (2019) .

\bibitem{Cohen:2008wz}
T.~D. Cohen and D.~A. McGady, ``{The Schwinger mechanism revisited},''
  \href{http://dx.doi.org/10.1103/PhysRevD.78.036008}{{\em Phys. Rev. D}
  {\bfseries 78} (2008) 036008},
  \href{http://arxiv.org/abs/0807.1117}{{\ttfamily arXiv:0807.1117 [hep-ph]}}.

\bibitem{nachman2020unfolding}
B.~Nachman, M.~Urbanek, W.~A. de~Jong, and C.~W. Bauer, ``Unfolding quantum
  computer readout noise,'' {\em npj Quantum Information} {\bfseries 6} no.~1,
  (2020) 1--7.

\bibitem{he2020resource}
A.~He, B.~Nachman, W.~A. de~Jong, and C.~W. Bauer, ``Resource efficient zero
  noise extrapolation with identity insertions,'' {\em arXiv preprint
  arXiv:2003.04941} (2020) .

\bibitem{Schwinger:1962tp}
J.~S. Schwinger, ``{Gauge Invariance and Mass. 2.},''
  \href{http://dx.doi.org/10.1103/PhysRev.128.2425}{{\em Phys. Rev.} {\bfseries
  128} (1962) 2425--2429}.

\bibitem{Lowenstein:1971fc}
J.~H. Lowenstein and J.~A. Swieca, ``{Quantum electrodynamics in
  two-dimensions},'' \href{http://dx.doi.org/10.1016/0003-4916(71)90246-6}{{\em
  Annals Phys.} {\bfseries 68} (1971) 172--195}.

\end{thebibliography}\endgroup

\newpage

\onecolumngrid

\fontsize{12pt}{14pt}\selectfont
\setlength{\parindent}{15pt}
\setlength{\parskip}{1em}

\newpage

\begin{center}
	\textbf{\large 3+1 Dimension Schwinger Pair Production with Quantum Computers} \\ 
	\vspace{0.05in}
	{ \it \large Supplemental Material}\\ 
\end{center}
\vspace{0.05in}
\appendix
\section{Theoretical setup}
\subsection{Continuum formulation, (3+1)D to (1+1)D}
In this section we show that the Schwinger pair-production in an electric background field in (3+1)D can be decomposed into an (1+1)D QED problem (i.e. the Schwinger model\cite{Schwinger:1962tp, Lowenstein:1971fc}).

First, we define the (3+1)D Dirac field in the Schr\"odinger picture:
\begin{align}
	\nonumber\psi(\vec x)&=\int \frac{d^3p}{(2\pi)^3}\frac{1}{\sqrt{2\omega_p}}e^{i\vec p\cdot\vec x}\sum_s(a_{\vec p}^s u^s(\vec p)+b_{-\vec p}^s v^s(-\vec p))\\
	&=\int\frac{dp_xdp_y}{(2\pi)^2}e^{i(p_xx+p_yy)}\sum_s \psi_s(z,p_x,p_y),
\end{align}
where 
\begin{equation}
	\psi_s(z,p_x,p_y)=\int \frac{dp_z}{2\pi}\frac{1}{\sqrt{2\omega_p}}e^{i p_zz}(a_{\vec p}^s u^s(\vec p)+b_{-\vec p}^s v^s(-\vec p)).
\end{equation}

The free Hamiltonian is of the form
\begin{equation}
	H=\int d^3x\psi^\dagger(-i\gamma^0\vec\gamma\cdot\nabla+m\gamma^0)\psi=\int dz\int\frac{dp_xdp_y}{(2\pi)^2}\sum_s\psi_s^\dagger h \psi_s,
\end{equation}
where gamma matrices are in the Dirac basis
\begin{equation}
	\gamma^0=\begin{pmatrix}I_{2\times 2} & 0\\ 0 & -I_{2\times 2} \end{pmatrix},\quad
	\gamma^i=\begin{pmatrix}0 & \sigma^i\\ -\sigma^i & 0\end{pmatrix},
\end{equation}
and
\begin{equation}
	h=\begin{pmatrix}
		m & 0 & -i\partial_z & p_x-i p_y\\
		0 & m & p_x+i p_y & i\partial_z\\
		-i\partial_z & p_x-i p_y & -m & 0\\
		p_x+i p_y & i\partial_z & 0 & -m
	\end{pmatrix}.
\end{equation}

By doing the unitary transformation
\begin{equation}
	U=\frac{1}{\sqrt{2m'}}\begin{pmatrix}
		\sqrt{m'+m} & -\frac{p_x\sigma_x+p_y\sigma_y}{\sqrt{m'+m}}\\
		\frac{p_x\sigma_x+p_y\sigma_y}{\sqrt{m'+m}} & \sqrt{m'+m}
	\end{pmatrix},
\end{equation}
we obtain
\begin{equation}
	h'=U^\dagger h U=
	\begin{pmatrix}m' I_{2\times2} & -i\partial_z\sigma_z\\
		-i\partial_z\sigma_z & -m' I_{2\times2}\end{pmatrix},
\end{equation}
where $m'=\sqrt{m^2+p_x^2+p_y^2}$ and we define $\tilde \psi_s=U^\dagger\psi_s$.

One can easily find the eigenvectors of  $h'$, which are also the solutions to the free-particle Dirac eqation, to be
\begin{equation}
	u'_{\frac{1}{2}}(\vec p)\propto \begin{pmatrix} m'+\omega_p\\0\\p_z\\0 \end{pmatrix},\quad
	v'_{\frac{1}{2}}(\vec p)\propto \begin{pmatrix} m'+\omega_p\\0\\-p_z\\0 \end{pmatrix},\quad
	u'_{-\frac{1}{2}}(\vec p)\propto \begin{pmatrix} 0\\m'+\omega_p\\0\\-p_z \end{pmatrix},\quad
	v'_{-\frac{1}{2}}(\vec p)\propto \begin{pmatrix} 0\\m'+\omega_p\\0\\p_z \end{pmatrix},
\end{equation}
which means the 2,4-components of $\tilde \psi_{1/2}$ and the 1,3-components of $\tilde \psi_{-1/2}$ are always zero. Thus it is natural to define the following two-component spin states
\begin{equation}
	\tilde\psi^{(2)}_{1/2}=\begin{pmatrix}\tilde\psi_1\\\tilde\psi_3\end{pmatrix}, \tilde\psi^{(2)}_{-1/2}=\begin{pmatrix}\tilde\psi_2\\\tilde\psi_4\end{pmatrix}.
\end{equation}

Now the Hamiltonian can be written as
\begin{equation}
	H=\sum_s\int\frac{dp_xdp_y}{(2\pi)^2}\tilde H^{(2)}_s,
\end{equation}
where
\begin{equation}\label{H_s}
	\tilde H^{(2)}_s=\int dz \tilde\psi^{(2)\dagger}_s(\pm i \partial_z\tilde\gamma^0\tilde\gamma^1+m'\tilde\gamma^0)\tilde\psi^{(2)}_s
\end{equation}
has exactly the same form of the Hamiltonian in (1+1)D, and 
\begin{equation}
	\tilde\gamma^0=\begin{pmatrix}1 & 0\\ 0 & -1 \end{pmatrix},\quad
	\tilde\gamma^1=\begin{pmatrix}0 & -1\\ 1 & 0\end{pmatrix}
\end{equation}
are the gamma matrices in 2D, the sign ambiguity in \cref{H_s} can be absorbed by redefining $\tilde\gamma^1$. 

Given a static electric background field along z-axis, choosing the axial gauge $\vec A=0, A_0=A_0(z)$, we need to add the following interaction term into the Hamiltonian
\begin{equation}
	H_I=\int d^3x e A_0\psi^\dagger\psi.
\end{equation}

Following the same procedure, the interaction term can be written as
\begin{equation}\label{H_I}
	H_I=\sum_s\int\frac{dp_xdp_y}{(2\pi)^2}\tilde H^{(2)}_{I,s},
\end{equation}
where 
\begin{equation}\label{H_sI}
	\tilde H^{(2)}_{I,s}=\int dz e A_0\tilde \psi_s^{(2)\dagger} \tilde\psi^{(2)}_s
\end{equation}
again exactly matches with the interaction term in 2D.

Now the (3+1)D QED problem are reduced to (1+1)D QED with the mass $m$ replaced by  $m'=\sqrt{m^2+p_x^2+p_y^2}$,  scanning over the momentum space along $x$ and  $y$ direction and summing over spin.

It can be shown that the vacuum decay rate of the Schwinger effect in (3+1)D are related to the rate in (1+1)D \cite{Cohen:2008wz}
\begin{equation}
	\Gamma_{3+1}(m)=2\int\frac{dp_xdp_y}{(2\pi)^2}\Gamma_{1+1}(m').
\end{equation}

\subsection{Lattice formulation of (1+1)D fermions}

Having shown that the Schwinger pair production of fermions with mass m and momentum $\vec p$ in (3+1)D is equivalent with the pair production of (1+1)D fermions with mass $m'=\sqrt{m^2+p_x^2+p_y^2}$ and momentum $p=p_z$, in this section we present the discretization of the (1+1)D Hamiltonian \cref{H_s,H_sI} on a lattice grid.
We place the fermion field on a 1D lattice with spacing $a$, labeled by $n$. To avoid the fermion doubling problem, we adopt the staggered fermion approach \cite{Kogut:1974ag, Banks:1975gq}, where the upper(lower) component of $\tilde\psi_s$ is mapped to the fermion field $\phi(n)$ at even (odd) lattice sites 
\begin{equation}
	\phi(n)/\sqrt{a}\rightarrow\begin{cases}
		\tilde\psi^{(2)}_{\text{upper}}(n a) & n \text{ even,}\\
		\tilde\psi^{(2)}_{\text{lower}}(n a) & n \text{ odd.}
	\end{cases}
\end{equation}

Fermion fields on each site satisfy the following anti-commutation relation
\begin{equation}
	\{\phi^\dagger(m),\phi(n)\}=\delta_{mn},\quad\{\phi(m),\phi(n)\}=0.
\end{equation}

The Hamiltonian \cref{H_s,H_sI} are then discretized to be
\begin{align}
	H_0&=\frac{i}{2a}\sum_n[\phi^\dagger(n)\phi(n+1)-\phi^\dagger(n+1)\phi(n)]+m\sum_n(-1)^n\phi^\dagger(n)\phi(n),\\
	H_I&=\sum_n e A_0(n a)\phi^\dagger(n)\phi(n).
\end{align}

We adopt $A_0(z)=- E |z|$ with origin mapped at the site $N/2$.
Observing that the Hamiltonian $H=H_0+H_I$ is invariant under the parity transformation $P:\tilde\psi^{(2)}_s(z)\rightarrow\tilde\gamma^0\tilde\psi^{(2)}_s(-z)$ for continuum 
or $P: \phi(m)\rightarrow(-1)^m\phi(N-m)$ for discrete, we could define the parity even and odd field
\begin{equation}
	\phi_\pm(m)=\frac{\phi(m)\pm (-1)^m\phi(N-m)}{\sqrt{2}},\quad m=1,2,\dots,\frac{N}{2}-1,
\end{equation}
\begin{equation}
	\phi_+(0)=\phi(0),\quad \phi_-(0)=0,\quad \phi_\pm(\frac{N}{2})=\frac{1\pm (-1)^{N/2}}{2}\phi(\frac{N}{2}).
\end{equation}
which satisfy the anti-commutation relations
\begin{equation}
	\{\phi^\dagger_\pm(m),\phi_\pm(n)\}=\delta_\pm\delta_{mn}.
\end{equation}

The Hamiltonian is further divided into two parts $H=H_++H_-$ which can be simulated separately 
\begin{align}
	H_+&=\frac{i}{2a}\sum_{m=1}^{N/2-2}[\phi^\dagger(m)\phi(m+1)-\phi^\dagger(m+1)\phi(m)]\nonumber\\
	&+\frac{i}{\sqrt{2}a}[\phi^\dagger(0)\phi(1)-\phi^\dagger(1)\phi(0)]\nonumber\\
	&+\sum_{m=0}^{N/2-1}[(-1)^m M+e E a m]\phi^\dagger(m)\phi(m),\\
	H_-&=\frac{i}{2a}\sum_{m=1}^{N/2-2}[\phi^\dagger(m)\phi(m+1)-\phi^\dagger(m+1)\phi(m)]\nonumber\\
	&+\frac{i}{\sqrt{2}a}[\phi^\dagger(\frac{N}{2}-1)\phi(\frac{N}{2})-\phi^\dagger(\frac{N}{2})\phi(\frac{N}{2}-1)]\nonumber\\
	&+\sum_{m=1}^{N/2}[(-1)^m M+e E a m]\phi^\dagger(m)\phi(m).
\end{align}

The fermion field operators can be written in terms of spin operators by applying the Jordan-Winger transformation \cite{jordan1993paulische}
\begin{align}
	\phi(n)&=\prod_{l<n}[i\sigma_3(l)]\sigma^-(n),\\
	\phi^\dagger(n)&=\prod_{l<n}[-i\sigma_3(l)]\sigma^+(n),
\end{align}
where $\sigma^{\pm}(n)=[\sigma_1(n)\pm i\sigma_2(n)]/2$.

The Hamiltonian is then converted to
\begin{align}
	H_+&=\frac{1}{2a}\sum_{n=1}^{N/2-2}[\sigma^+(n)\sigma^-(n+1)+\sigma^+(n+1)\sigma^-(n)]\nonumber+\frac{1}{\sqrt{2}a}[\sigma^+(0)\sigma^-(1)+\sigma^+(1)\sigma^-(0)]\nonumber\\\label{H_+}
	&+\sum_{n=0}^{N/2-1}[(-1)^n m+e E a n]\frac{\sigma_3(n)+1}{2},\\
	H_-&=\frac{1}{2a}\sum_{n=1}^{N/2-2}[\sigma^+(n)\sigma^-(n+1)+\sigma^+(n+1)\sigma^-(n)]\nonumber+\frac{1}{\sqrt{2}a}[\sigma^+(\frac{N}{2}-1)\sigma^-(\frac{N}{2})+\sigma^+(\frac{N}{2})\sigma^-(\frac{N}{2}-1)]\nonumber\\
	&+\sum_{n=1}^{N/2}[(-1)^n m+e E a n]\frac{\sigma_3(n)+1}{2}.
\end{align}
which is ready to be simulated on a digital quantum computer.

\section{Demonstration of quantum circuits}
We decompose the parity even Hamiltonian into 3 pieces $H_+=H_{+1}+H_{+2}+H_{+3}$
\begin{align}
	H_{+1}&=\sum_{n=0}^{N/2-1}[(-1)^n m+e E a n]\frac{\sigma_3(n)}{2},\\
	H_{+2}&=\frac{1}{\sqrt{2}a}[\sigma^+(0)\sigma^-(1)+\sigma^+(1)\sigma^-(0)]  +\frac{1}{2a}\sum_{n=2}^{N/2-3}[\sigma^+(n)\sigma^-(n+1)+\sigma^+(n+1)\sigma^-(n)],\\
	H_{+3}&=\frac{1}{2a}\sum_{n=1}^{N/2-2}[\sigma^+(n)\sigma^-(n+1)+\sigma^+(n+1)\sigma^-(n)],
\end{align}
on which we could apply Suzuki-Trotter formula and simulate the time evolution. 

The quantum circuit for each Trotter step are shown below ($N/2=5$)
\begin{center}
	\includegraphics[width=0.5\textwidth]{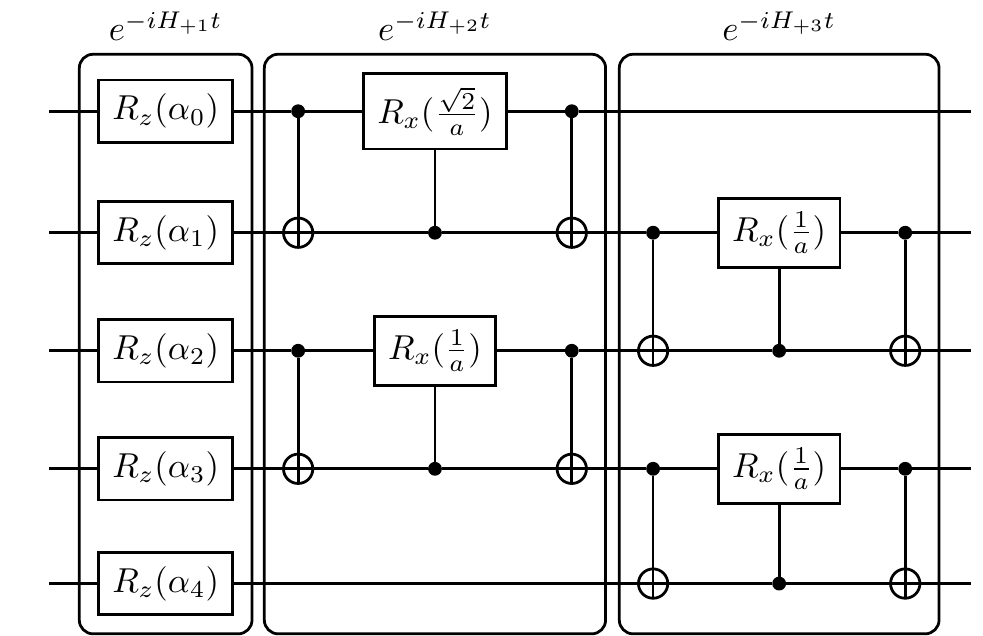}
\end{center}
where $\alpha_k=(-1)^k m + e E a k$.
Single qubit gates
\begin{equation}
	R_P(\theta)=e^{-i\frac{\sigma_P}{2}\theta}, \text{ where }P=x,y,z,
\end{equation}
represent rotations along the axis-P of the Bloch sphere.

The exponential of the operator $\sigma^+(n)\sigma^-(n+1)+\sigma^+(n+1)\sigma^-(n)$ in the kinetic term can be implemented by the following gates
\begin{center}
	\includegraphics[width=0.75\textwidth]{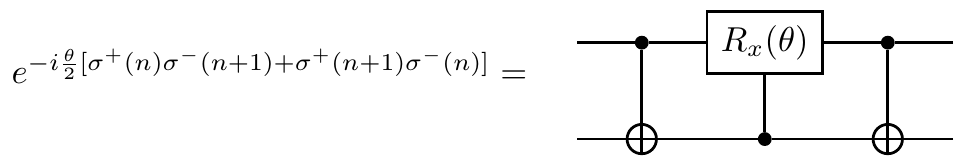}
\end{center}
where a controlled-x rotation is sandwiched by two CNOT gates, and is further transplied into basic gates of {\it ibm\_lagos} in the following structure with two CNOT gates
\begin{center}
	\includegraphics[width=0.75\textwidth]{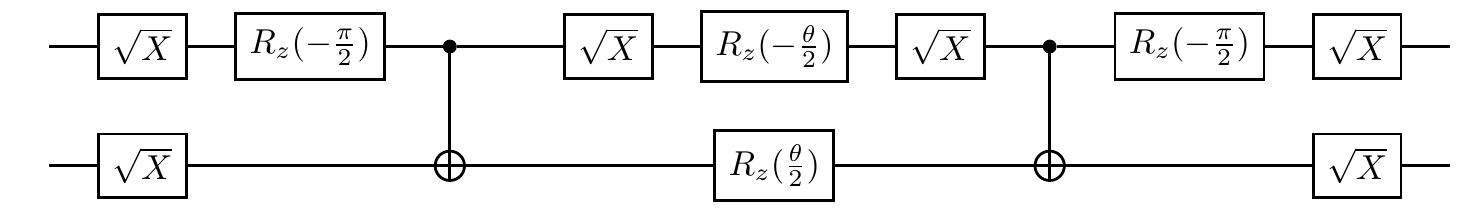}
\end{center}

The VQE ansatz we used to generate the ground state for $H_0$ is the following circuit with 9 parameters
\begin{center}
	\includegraphics[width=0.5\textwidth]{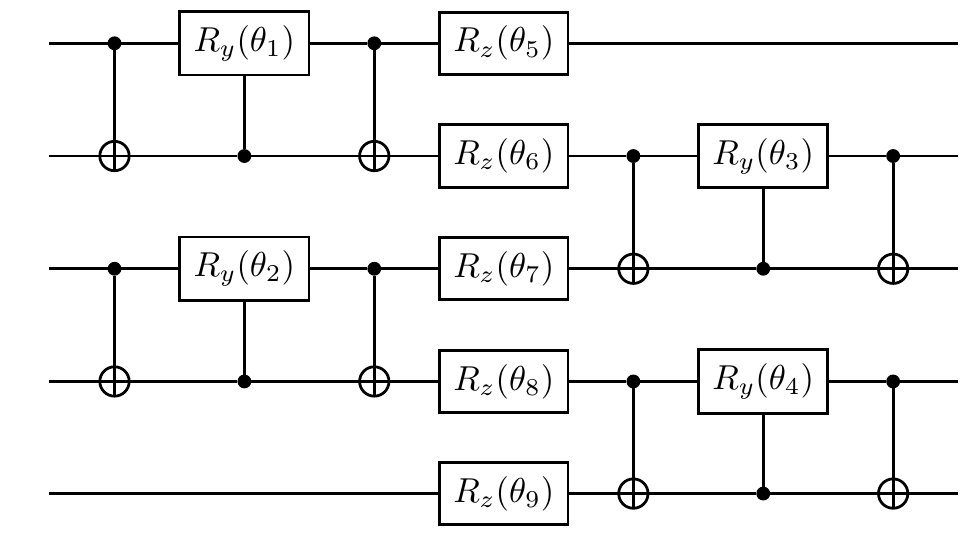}
\end{center}
where use controlled-y rotations sandwiched by two CNOT gates to generate entanglements between adjacent qubits. This operation is charge conserving and the corresponding unitary representations are all real. We then insert five z-rotations in the middle to generate phases.

\newpage

\section{Plots of vacuum persistence probability for different masses}

\begin{figure}[!h]
	\centering
	\includegraphics[width=\textwidth]{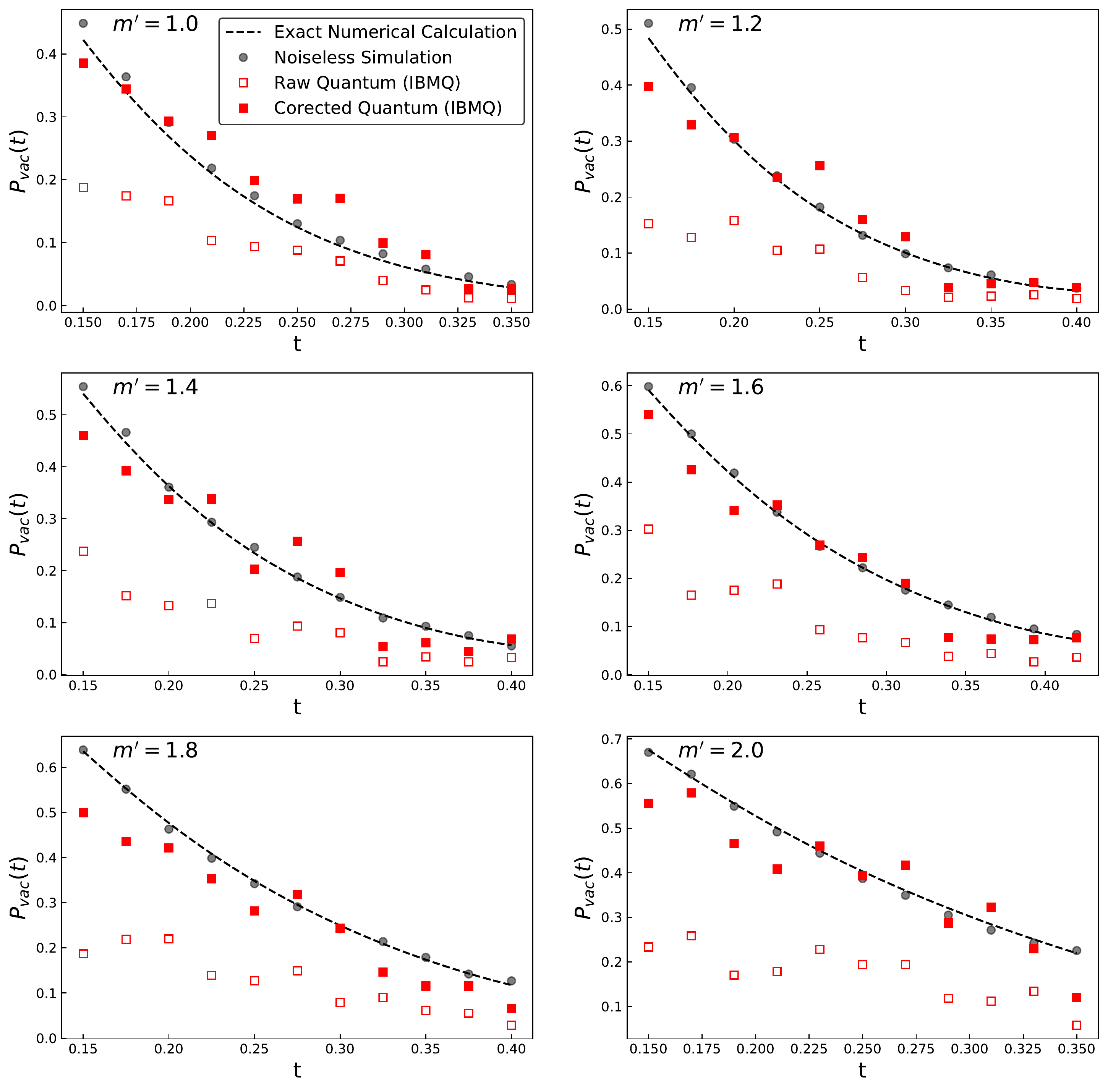}
	\caption{Vacuum persistent probability for the effective masses $m' = 1.0, 1.2, 1.4, 1.6, 1.8\text{ and } 2.0$.
		The dashed curve corresponds to the exact numerical solution of the Hamiltonian. 
		The gray dots show the simulation results on a noiseless simulator. 
		The hollow squares show results from the {\it ibm\_lagos} quantum computer. 
		The solid squares are the corrected results to the quantum simulation by restricting to physical subspace.
		The time regions we choose are $0.15<t<0.35$ for $m=1.0 \text{ and } 2.0$, $0.15<t<0.40$ for $m=1.2, 1,4 \text{ and }1.8$,  $0.15<t<0.42$ for $m=1.6$.}
\end{figure}

\end{document}